\documentclass{aa}
\usepackage{epsfig}
\usepackage{graphicx}
\begin{document}

\title{Searching for the in-plane Galactic bar and ring in DENIS}

\author{M. L\'opez-Corredoira\inst{1} \and P. L. Hammersley\inst{1} \and 
F. Garz\'on\inst{1,2} \and A. Cabrera-Lavers\inst{1}
\and N. Castro-Rodr\'\i guez\inst{1} \and M. Schultheis\inst{3} 
\and T. J. Mahoney\inst{1}}
\institute{Instituto de Astrof\'\i sica de Canarias, E-38200 La Laguna, Tenerife, Spain \and
Departamento de Astrof\'\i sica, Universidad de La Laguna, Tenerife, Spain
\and Institut d'Astrophysique de Paris, F-75014 Paris, France}

\offprints{martinlc@ll.iac.es}

\date{Received xxxx / Accepted xxxx}

\abstract{New evidence for a long thin Galactic bar (in contradistinction to the bulge),
as well as for the existence of the ring and the truncation of the inner disc,
are sought in the DENIS survey.
First, we examine  DENIS and Two Micron Galactic Survey  star counts for the characteristic 
signatures of an in-plane bar and ring.
The star counts in the plane for
$30^\circ>l>-30^\circ$ are shown to be highly asymmetric with
considerably more sources at positive than at negative
longitudes. 
At $|b|\approx 1.5^\circ$, however, the counts are nearly symmetric.
Therefore, the asymmetry is not due to the disc, which is shown to
have an inner truncation, or to the  bulge, so  there has to be another major component in the 
inner Galaxy that is causing the asymmetries. This component provides up to
50\% of the detected sources in the plane between the bulge and
$l=27^\circ$ or $l=-14^\circ$.  This 
component is shown to be consistent with an in-plane
bar with a position angle of $40^\circ$ and half-length of 3.9 kpc.
However, there is also a major peak in the counts at $l=-22^\circ$, which
coincides with the tangential point of the so-called 3 kpc arm. This
is shown to be most probably a ring or a pseudo-ring. 
The extinction in the
plane is also shown to be asymmetric with more extinction at negative
than at positive longitudes. For $l<8^\circ$ the extinction is shown to be
slightly tilted with respect to $b=0^\circ$  in the same manner as the HI
disc. 
 We conclude that the Galaxy is a fairly  
typical ringed barred spiral galaxy.
}
\titlerunning{The in-plane Galactic bar and ring}

\maketitle

\section{Introduction}

De Vaucouleurs (1964, 1970) first suggested, 
in an attempt to explain observed non-circular gas
orbits, that the Galaxy might be barred. Since then, a large body of observational evidence
has been accumulated that supports this hypothesis
(see Garz\'on 1999 for a review). 
Axial asymmetries in the inner Galaxy have been detected in  
star counts (Nakada et al. 1991; Weinberg 1992; 
Whitelock et al. 1992; Stanek et al. 1994; Hammersley et al. 1994, 1999, 2000
W\'ozniak \& Stanek 1996; Nikolaev \& Weinberg 1997; Unavane \& Gilmore
1998; L\'opez-Corredoira et al. 1997, 2000) and
by surface photometry at different wavelengths
(Blitz \& Spergel 1991a; Weiland et al. 1994; Dwek et al. 1995; Freudenreich
1998), microlensing (Stanek 1995; Binney et al. 2000) and
analysis of internal motions of the gas
(Peters 1975; Liszt \& Burton 1980; Yuan 1984; Nakai 1992; Gerhard 1996). 
Various models have been constructed to explain the observed features, many of 
which include a bar (e.g., Binney et al. 1991; Weiner \& Sellwood 
1999). The observed non-axisymmetry, however, varies considerably between 
the various papers, although ``bar'' is the term normally used. 
Unfortunately, this terminology is ambiguous  
and has led to much confusion and controversy, 
as has been commented on in several papers (Kuijken 1996; Ng 1998;
L\'opez-Corredoira et al. 1999). This  is more
than a mere question of words, however. Many authors have classed what is 
a short fat structure as a bar, when possibly a more appropriate term would 
be a triaxial bulge. Other authors have found a long thin structure, which,
if it exists,   would be what is traditionally considered to be a bar, 
and such a feature is inherently different from a triaxial bulge. 
 
Whilst evidence for the triaxial bulge is now overwhelming and the
results are basically consistent, there is still discussion on the
presence of the long thin bar. Radio maps show significant
non-axisymmetry in the motions of the gas in the inner galaxy. Peters
(1975) analysed the HI maps and showed that a bar 
inclined at about 45$^\circ$ to the Sun--Galactic Centre line would
produce the features seen. Nakai (1992) found a similar angle when
analysing CO maps, and a rough calculation places the ends of the bar at
$l\sim 30^\circ$ and $l=-20^\circ$.

{\it IRAS} star counts in the plane have a very asymmetric distribution
in longitude (e.g. Garz\'on 1999), with more stars at positive  than at
negative longitudes. These asymmetries stretch from about $l=30^\circ$
to $-30^\circ$ and are far larger than the asymmetries caused by a
triaxial bulge,  in both extent and magnitude.  Weinberg (1992) analysed
the {\it IRAS} star counts in the plane and proposed that there was an
in-plane stellar bar with a position angle of 36 $\pm 10^\circ$ and
half-length of about 5 kpc. Hammersley et al. (1994, hereafter H94)
examined data from the Two Micron Galactic Survey (TMGS) and showed
that there is a high density of young  stars in the plane at
$l=27^\circ$ and $21^\circ$ that is not seen either towards the bulge
or at  longitudes greater $l=27^\circ$.  An analysis of the {\em
COBE}/DIRBE 2.2-$\mu$m surface brightness maps showed that the form was
consistent with an in-plane bar.

Further evidence for the bar at positive longitudes is presented by
Garz\'on et al. (1997) and L\'opez-Corredoira et al. (1999),who made a
spectroscopic analysis of the brightest stars in an infrared-selected sample
of objects close to the Galactic plane at $l=27^\circ $ showing a strikingly
high fraction of supergiants, characteristic of a strong star formation region.
The typical distance to these sources was found to be 6 kpc.
It was argued that this result is consistent with this region being 
the near end  of the Galactic bar.
Such regions can form due to the concentrations of
shocked gas where the  stellar bar meets the disc, as is observed at
the ends of the bars of face-on external galaxies (Sandage \& Bedke 1994).
Hammersley et al. (2000, hereafter H00) 
examined infrared colour--magnitude 
diagrams in a number of regions in the plane. They show that there is a major 
cluster of K--M giants at a distance of 5.7 $\pm$ 0.5 kpc at $l=27^\circ$, 
which is not present at $l=32^\circ$. This cluster is seen at most
longitudes smaller than $l=27^\circ$ (the exception being regions of
extremely high extinction) and the distance from the Sun to the
cluster increases with decreasing longitude until about $l=10^\circ$, where it
merges with the bulge. H00 argue that what is being seen at
$l=27^\circ$ is the older component at the near end of a bar with a
position angle of 43$\pm 7^\circ$ and half length about 4 kpc. This would
put the far end of the bar near $l=-12^\circ$, and the authors note
that there is a peak in the {\em COBE}/DIRBE surface brightness maps at this
location. 

Other authors, however, prefer to explain these extra stars  seen in
the plane as belonging to a ring or spiral arms. Kent et al. (1991)
suggested a model containing a thick ring and disc which did
successfully reproduce the surface brightness maps at positive
longitudes (their data set was principally for positive longitudes).
However, their model is basically symmetrical and so would not
reproduce the negative longitudes as seen in {\em COBE}/DIRBE. Even
making the ring elliptical would not significantly improve the fit. In fact
none of the proposed models containing a bulge (including its
triaxiality) disc and ring successfully reproduces the star counts or
surface brightness maps in the plane between $l=40^\circ$ and
$-40^\circ$. Freudenreich (1998) has significant residuals in the plane
between $l=30^\circ$ and $-30^\circ$ after subtracting the disc and
bulge. He prefers to attribute these residuals to patchy star formation
trailing from a ring or spiral arm, however the discovery of older
stars in H00 rules out  patchy star formation; furthermore,
the residuals are not ``patchy'' but almost constant between $l=30^\circ$
and $-12^\circ$.

Weiner \& Sellwood (1999) produced a gravitational
model that included a separate bar and bulge, and with
this model they fitted the observed features caused by
the gas kinematics in the inner regions of
Milky Way. Their bulge is axisymmetric and the inclination of the bar
is $\sim 35^\circ $. In that paper, the authors claim that their
treatment of bulge and bar separately does not imply that both
components are different, either photometrically or kinematically.
However, although they do not make such a claim, it is evident that the
model works much better by introducing these two types of density
distributions.  Even if they were part of the same structure, 
the central kiloparsec would be significantly different from the  outer regions.

Generally, the observations indicate that size and form of the 
asymmetry in the star distributions  vary with  Galactic latitude.
Normally, when the low latitudes are excluded (i.e. observing the triaxial bulge) 
an angle of about 25$^\circ$ is found.  When the plane regions are included, however, 
the angle  reaches 45$^\circ$ (Sevenster et al. 1999). The 
difference in counts between positive and negative longitudes
is also larger close to the plane but it can still be detected  
ten or more degrees from the plane. This arrangement is consistent with  
a triaxial bulge with an inclination
around 20$^\circ$ or less, dominating the off-plane regions, 
and a thin  bar with an  inclination around 45$^\circ$, which 
gives strong asymmetry only in in-plane positions. The change in angle 
with latitude would be related to the ratio of bulge and bar counts,
with each having different position angles.

An important parameter in any 
barred galaxy is the location of corotation. Corotation is at 1.1 to 1.2 
times the length of the bar and the bar cannot extend beyond corotation. 
However, a triaxial or boxy--peanut bulge will end near the vertical ILR, 
which is at about $R_{\rm cr}/2.2$ (Friedli 1999). Therefore, both the thin bar and 
triaxial bulge put corotation at about 4.4 to 5 kpc. This is in agreement 
with the conclusions of Combes (1996). Galaxies with two, or even 
three, triaxial structures are known (see Friedli 1996 for a review) 
and the position angles of the various components can take any value
(in fact, they are aligned only in a
minority of cases). Hence, it is quite reasonable that a
triaxial bulge and in-plane bar have slightly different position angles.

\begin{figure}
\begin{center}
\vspace{1cm}
\mbox{\epsfig{file=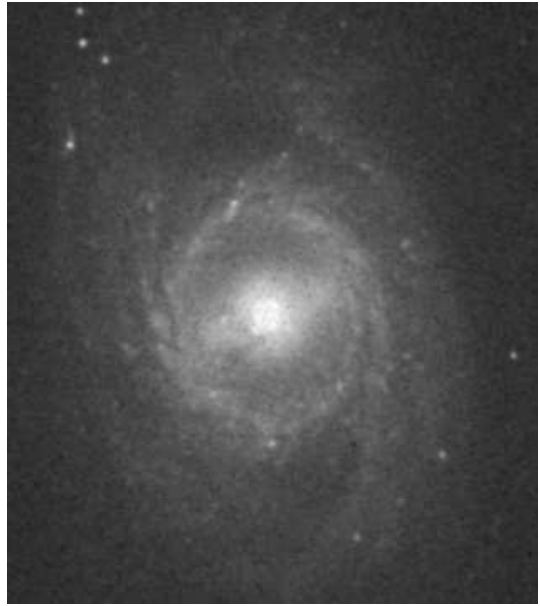,height=8cm}}
\end{center}
\caption{POSSII(J-Blue) image of the spiral galaxy M95 (NGC 3351). Note the 
presence of a bar apart from the prominent bulge in the centre of the
galaxy.}
\label{Fig:M95}
\end{figure}

\begin{figure}
\begin{center}
\vspace{1cm}
\mbox{\epsfig{file=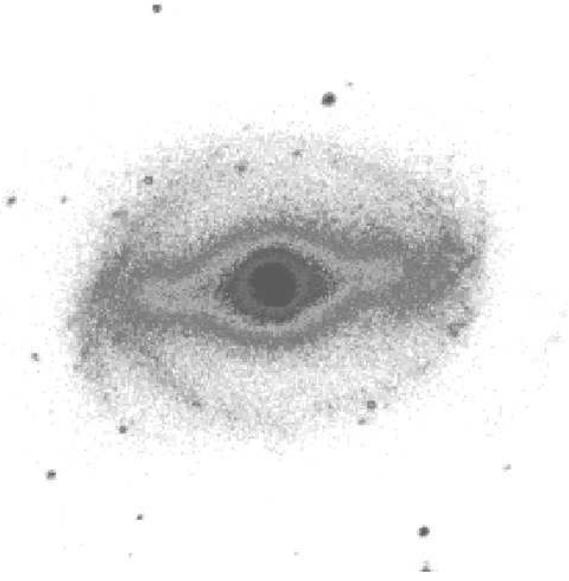,height=8cm}}
\end{center}
\caption{POSSII(R-Red) image of the spiral galaxy NGC 1433. Another
example of a barred galaxy with bulge}
\label{Fig:NGC1433}
\end{figure}

\begin{figure}
\begin{center}
\vspace{1cm}
\mbox{\epsfig{file=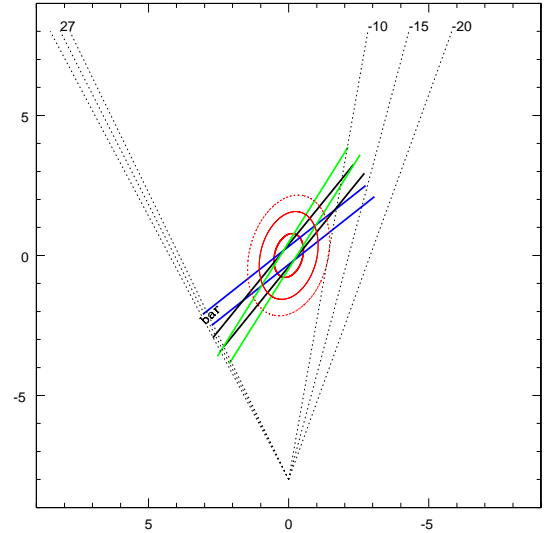,height=8cm}}
\end{center}
\caption{A Schematic representation of the 
proposed ``triaxial bulge'' + ``bar''
structures in the inner Galaxy. The bar width is assumed to be 500 pc. 
Three possible bars are plotted  showing the range of position angles 
determined 
from H00 and assuming that the near end of the bar is at $l=27^\circ$. 
The lines of sight for 
$l=26^\circ $, $27^\circ $, $28^\circ $, $-10^\circ $, 
$-15^\circ $, $-20^\circ $ are drawn. The Sun is at (0,$-$8) kpc.}
\label{Fig:galmoda}
\end{figure}

From the above it is clear that the non-axisymmetry in the inner Galaxy
is best described by two separate components: a ``triaxial bulge'' and a
``thin bar'' and the two should not be confused. The bar is long and thin,
and contains both young and older stars, whereas the triaxial bulge is
shorter, far fatter and only contains old stars (H94, H00, Garz\'on et
al. 1997, L\'opez-Corredoira et al. 2000).
This implies that the inner Galaxy when viewed face on would 
appear similar to M95 (NGC 3351), of revised Hubble type
SBb(r)II (Sandage \& Tammann 1981)
(Figure \ref{Fig:M95}), although the bulge of the 
Milky Way may be less prominent, or NGC 1433 (Fig.
\ref{Fig:NGC1433}), revised Hubble type SBb(s)I--II (Sandage
\& Tammann 1981), where ``r"" denotes ``ringed'' and ``s'' denotes that
the spiral arms originate either from the ends of the bar or the from the
centre, and the roman numerals I and II denote early type
and intermediate luminosities. Other authors interpret NGC 1433
in terms of a double-barred galaxy (Friedli 1999) with different 
assumptions (they assume a elliptical shape for the thin bar 
instead of a sticklike shape);
in any case, we have two different non-axisymmetric structures.
Figure \ref{Fig:galmoda} shows a graphical representation of the proposed 
configuration for the inner Galaxy. It includes:

\begin{description}
\item[A long, thin bar] (half-length 3.9 kpc) with the near end at
$l=27^\circ $. Three possible values for the angle between 
the bar and Sun--Galactic Centre line are shown in Fig. \ref{Fig:galmoda}: 
50$^\circ$, 43$^\circ$ and 36$^\circ$.
Such a bar would have star formation regions at its tips.

\item[A triaxial bulge,] which is shorter, far thicker (extending to $|b| \approx 
10-15^\circ $), with a typical angle of about 12$^\circ$ with respect to
the Sun--Galactic centre line, and containing only old population stars
(except for a small region in the inner bulge, L\'opez-Corredoira et al. 2001a).
\end{description}

The aim of this paper is to examine  DENIS star counts for evidence of the
long, thin bar and/or ring. Both the proposed continuous ring and bar
 make clear 
predictions as to the form of the star counts in the inner Galaxy, which 
can be tested against the data. In this paper, the term ``bar'' will  refer
only to the in-plane bar described in H00. When discussing the non-axisymmetry 
of the bulge, the term ``triaxial bulge'' will be used. 
The evidence for  asymmetry in in-plane regions is explored in
this paper through the analysis of star counts (\S \ref{.counts}) and 
extinction (\S \ref{.ext}).

\section{DENIS data}

The Deep Near-Infrared Survey of the southern sky (DENIS) is a survey
the southern hemisphere in three bands: Gunn $I$ (0.82 $\mu $m),
$J$ (1.25 $\mu $m), $K_{\rm s}$ (2.15 $\mu $m) with limiting
magnitudes 18.5, 16.5 and 14.0  and saturation magnitudes
9.5, 8.5 and 6.5, respectively (Epchtein 1998).
The ESO 1 m Telescope at La Silla (Chile) was used for the survey.
The observations were begun in 1995 December and will be 
completed by the end of 2001 approximately. Part of these
data are already reduced and a small part has already been released at the
CDS for public use (Epchtein et al. 1999). 
A special effort has been made to
cover priority regions of high scientific interest, such as the inner Galactic 
plane and bulge, and it is these data that will be used in this paper.
In total, an area of around 170 deg$^2$ within the area $|l|<35^\circ $, 
$|b|<2^\circ $ has been covered. As the positive longitudes
were less sampled than the negative longitudes
(96 deg$^2$ are available
at $-35^\circ <l<0^\circ $, $|b|<1.5^\circ $, while for positive
longitudes only half of the area is covered: 47 deg$^2$ are available
at $0^\circ <l<35^\circ $, $|b|<1.5^\circ $) some data from TMGS
(Garz\'on et al. 1993, 1996)
were used to complete the counts in those regions without DENIS data,
and to calculate the star counts of the section \S \ref{.counts}.
With regard to the extinction shown in \S \ref{.ext}, the calculations
were carried out from DENIS data by Schultheis et al. 
(private communication).

The images were reduced at the Paris Data Analysis Centre (PDAC) of DENIS,
and homogeneous criteria of reduction were applied to the whole special
sample of the bulge, which is used here. 
They were checked and calibrated (Borsenberger 1997; Ruphy et al. 1997)
at PDAC. The point source extraction was also carried out at PDAC by fitting PSFs 
optimized for crowded fields. For the astrometry, the individual DENIS frames 
were cross-correlated with the PMM catalogue (USNO-A2.0) with a precision
of better
than one arcsecond.

\section{Asymmetry in the star counts of the Galactic plane}
\label{.counts}

Star counts provide a powerful tool when searching for asymmetries in the
stellar 
distribution in the inner Galaxy. Other researchers
have already used this method (see introduction); 
however, they have mostly observed regions where the bulge
is predominant. While the bulge is observed in off-plane
regions up to $b\sim 10$  degrees and in $|l|< 15$ degrees, the long, thin
bar is visible only in the in-plane regions, $|b|<2$ degrees and up to
$|l|=27^\circ $ at positive longitudes, and somewhat less at negative
longitudes.
It is now well known that the star counts in the bulge are
non-axisymmetric (e.g L\'opez-Corredoira et al. 2000); however, the 
larger-scale in-plane asymmetries are less well investigated.

The most appropriate DENIS filter for probing the inner Galaxy near plane is 
$K_{\rm s}$, since the effect of extinction is much lower than in the other filters.
However, the other filters can be used to provide information about
the population of the stars and the extinction.

Although the nominal $K_{\rm s}$ limiting magnitude of DENIS is around 
14.0 (Epchtein 1998); the presence of
 confusion implies that in the more crowded regions 
the completeness can be two or three magnitudes brighter, particularly 
towards the inner bulge (Unavane et al. 1998).   
For this study, however, only the brighter stars will be used, 
so confusion is not an issue.
Whilst it is often argued that the fainter the star counts the better, this is 
not necessarily so when observing the inner Galaxy. 
The ratio of inner Galaxy to local disc stars is more important.  
At magnitude 13 or 14 there will be more inner Galaxy sources than at 
magnitudes 8 to 10, 
but the numbers of local disc sources rises even more rapidly. 
Garz\'on et al. (1993) showed that the best contrast when looking at the 
bulge was at near $m_{\rm K}$ = +8,  and that at  fainter magnitudes the contrast 
became poorer. However, the number of stars detected falls rapidly much 
brighter than this (few if any bulge sources are brighter than $m_{\rm K}$ = 6, 
L00) and the statistical errors become important. 
For this paper star counts up to $m_{K_{\rm s}}=9$ will be used, as the 
inner-Galaxy-to-disc contrast is high, but there are still sufficient 
sources to provide good statistics.

\begin{figure*}
\begin{center}
\vspace{1cm}
\mbox{\epsfig{file=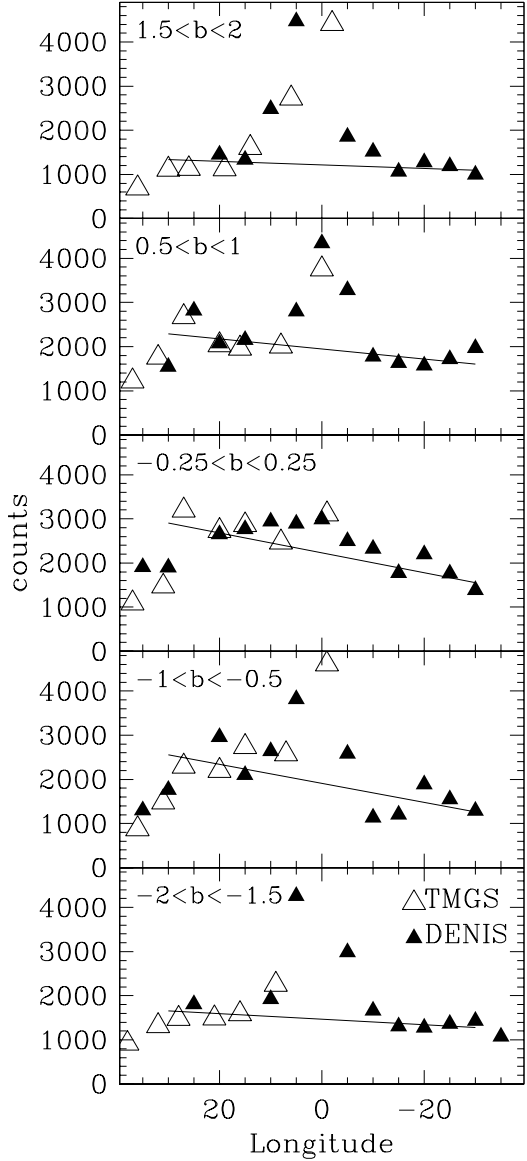,height=18cm}}
\end{center}
\caption{Star counts with 
$m_{K}\le 9.0$ at different Galactic latitudes. 
The crosses are data from the DENIS survey (Epchtein 1998) averaged over
$\Delta l=5^\circ$, and the circles are from TMGS (Garz\'on et al. 1993; 1996).
Large regions were used for each region with thousands of stars, so the
Poissonian error is negligible. Dashed lines show a
general trend in the star counts in $10^\circ <|l|<25^\circ $.}
\label{Fig:DENIScounts9}
\end{figure*}

The star counts down to $m_{K_{\rm s}}$ = 9.0 are shown in  Figure
\ref{Fig:DENIScounts9}. The DENIS star counts show the average of the
available  counts in a region of width $\Delta l=5^\circ $ centred
on the position marked. A large bin size was adopted, as this averages
out the fluctuation due to the varying extinction and the asymmetries
being sorted are distributed over many  tens of degrees. 
To complete the longitude
coverage, the star counts from TMGS are used (see Hammersley et al.
1999 for more details on the data). The TMGS data, however, are
plotted at the average longitude where the strip crossed the
plane. In both cases the areas of sky covered were large so that  the
Poissonian error is negligible. There is a difference between the
standard $K$ filter of the TMGS and the DENIS $K_{\rm s}$; 
however, Cohen (1997) has
shown that the difference between the two for any star is less than 0.03
mag, so the effect on the counts of using one or other should be
negligible.

Fig. \ref{Fig:DENIScounts9} shows five slices across the plane at
$2^\circ>b>1.5^\circ$, $ 1^\circ>b> 0.5^\circ$, $ 0.25^\circ>b>-0.25^\circ$, 
$-0.5^\circ>b>-1^\circ$ and $-1.5^\circ>b>-2^\circ$.
Therefore, at the distance of the Galactic Centre (8 kpc) the average heights
over the plane are 0 pc, $\pm 100$ pc  and $\pm$ 240 pc. It is
noticeable that the cuts furthest from the plane are substantially
different from those in the plane. It can further be seen that the
counts  are more or less symmetrical in latitude. The slight
differences that there are can be attributed to the Sun being about
14 pc above the Galactic plane (Hammersley et al. 1995),
which would give  more counts at negative 
 than at positive latitudes, and differences in the extinction above and 
below the plane. The small differences that there are between the TMGS and 
DENIS counts can be attributed to the fact that they are centred on
different longitudes and cover different areas of sky. 

\begin{figure}
\begin{center}
\vspace{1cm}
\mbox{\epsfig{file=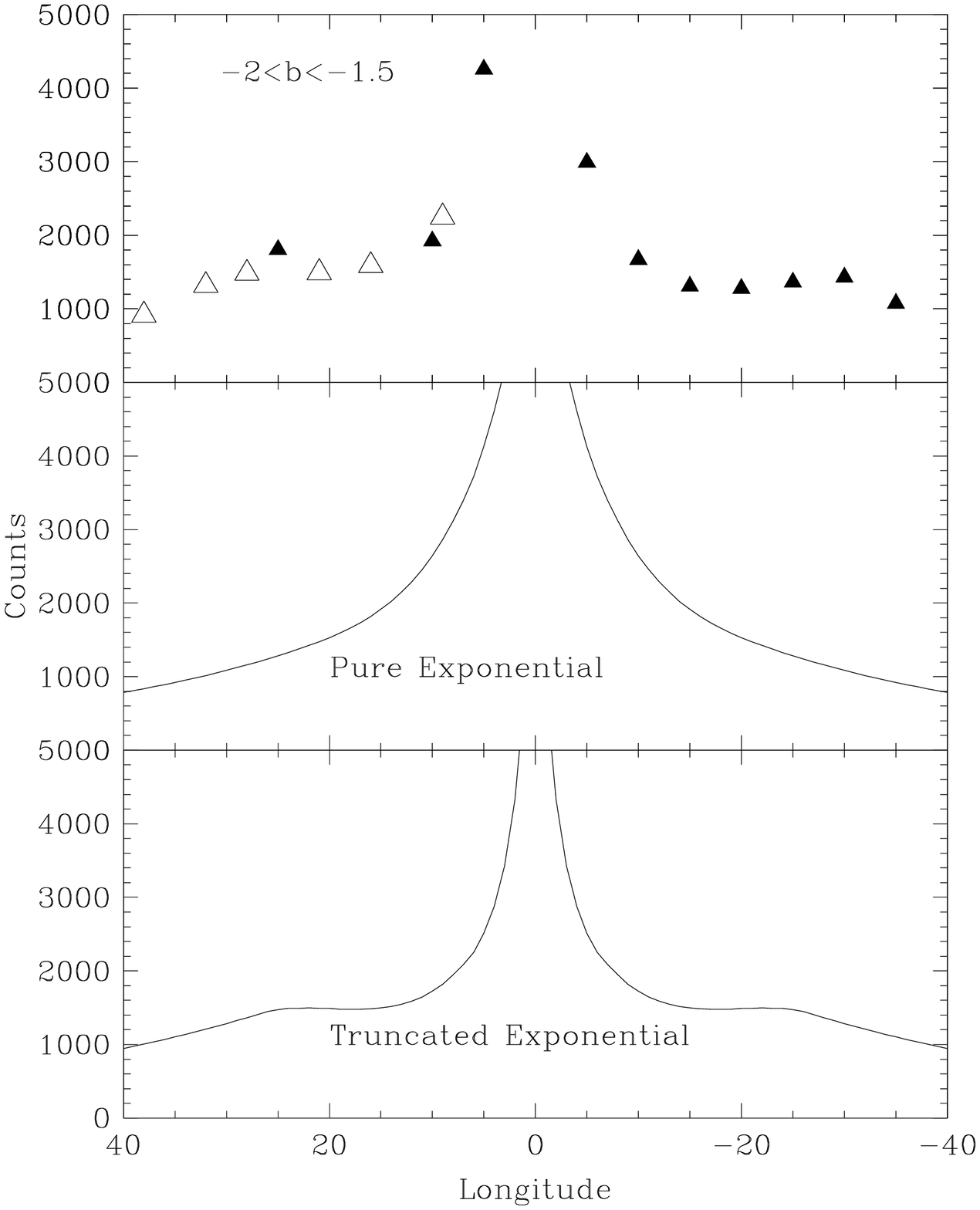,height=11.5cm}}
\end{center}
\caption{Star counts with 
$m_{K}\le 9.0$ for $-1.5^\circ >b>-2^\circ $ 
and the predicted counts using two simple disc--bulge models (see text).}
\label{Fig:offplan}
\end{figure}

\subsection{Off-plane star counts}
Before attempting to disentangle the  $b=0^\circ$ slice, it is important to understand the shape of the disc, as this will be an important feature in the plane. 
The slices at $|b|=1.75^\circ$ should be very sensitive to the shape
 of the disc.
They are far enough from the plane that the extinction will be low, typically
2 to 3 magnitudes in the visible or 0.2 to 0.3 mag at $K_{\rm s}$. 
Furthermore, the lines of sight run sufficiently far above and below the
plane that the young components close to the plane will not be a major
 component 
in the counts and only inner Galaxy features with scale heights in the 
hundreds of parsecs should provide a significant contribution.

Fig. \ref{Fig:DENIScounts9} shows that the slices  for $|b|=1.75^\circ$ are more or less symmetric in longitude. Initially, the counts rise with decreasing longitude
but then  for $30^\circ>|l|>12^\circ$ the counts are basically flat  before rising steeply for $|l|<10^\circ$ as the bulge counts become increasingly important. These plateaux
are not expected for a purely exponential disc.
Fig. \ref{Fig:offplan} shows the counts for $-1.5^\circ >b>-2^\circ $ 
and the predicted counts using two simple disc--bulge models. Both models
are based on Wainscoat et al. (1992), whose model 
has been shown to reproduce the TMGS star counts very well
(Hammersley et al. 1999). The bulge used
is that described in Wainscoat et al. (1992); the fact that the model is axisymmetric 
whereas the bulge is triaxial is not important for the following discussion. 
The extinction model used is also as described in Wainscoat et al. (1992). The
difference between the two plots is that the figure marked ``exponential
disc'' has an exponential disc that continues into the Galactic
Centre, again as in Wainscoat et al. (1992). However, the model marked ``truncated  exponential'' 
has a stellar density that only rises exponentially inwards down to 
3.5~kpc, from which point it falls off linearly such that at the centre it 
is zero. This approximates to a Freeman type II disc with a central hole and 
is known as an inner truncated disc.

The model predicts that a purely exponential disc that continues
into the centre should rise increasingly steeply  with decreasing
longitude; however, the measured counts clearly do not do this. The
truncated model does, however, reproduce the shape very well.  A full
discussion of the exact form of the inner disc is beyond the scope of
this paper, as data further from the plane are required; however, this
result is in agreement with the findings from the analysis of the {\em
COBE}/DIRBE surface brightness maps (e.g. Freudenreich 1998). Inner truncated discs
are very common in barred spirals. Ohta et al. (1990)
looked at six early-type  spirals and found that all had Freeman type
II discs when looking perpendicular to the 
plane. Baggett et al. (1996) 
show that barred galaxies are a factor of two, or
more, more likely than non-barred Galaxies to have an inner truncated
disc. They also note that an increasing number of bars is being found in
galaxies previously classified as non-barred, so this percentage is
likely to rise. 

\subsection{In-plane star counts}

The large-scale asymmetry counts in the plane are very noticeable. Positive
longitudes consistently have far more counts than negative longitudes. 
Furthermore, the shape of the counts is very different at positive and 
negative longitudes, so the form cannot be explained by simply making the 
inner Galaxy elliptical. From about $l=27^\circ$ to 
the Galactic Centre the counts are flat, whereas between $l=0^\circ$ and 
$l=-18^\circ$ they are reduced by a factor 2. This asymmetry is 
greater in the $b=0^\circ$ strip than in the $|b|=0.75^\circ$ strip 
because the relative contribution of the bar will be larger at $b=0^\circ$. 
The disc will  contribute about 50\% of the 
counts at $l=20^\circ$ in the plane but, as is shown in the 
previous section, the disc is symmetric. Hence, when the disc is subtracted 
from the in plane counts the asymmetries between positive and negative 
longitudes in the remaining counts (i.e. those from the inner Galaxy 
components) become enormous. 

The peak due to the bulge at $l=0^\circ$ is hardly seen at all in the
in-plane strip, whereas it is clearly evident in the off-plane counts.
This is due in part to the strong extinction within a few hundred pc
of the Galactic Centre (GC) which drastically reduces the counts in the
plane (Hammersley et al. 1999). Hence, the rapid increase in star
density near the GC is masked.  
However, the $l=7^\circ$ $b=0^\circ$
line of sight runs far enough away from the GC not to be
affected by the very high extinction near the GC, and here the bulge
contributes about 50\% of the counts to $m_K=9$ (Fig. B2 in Hammersley
et al. 1999). At greater absolute longitudes the bulge quickly dies away,
providing a negligible contribution for  $|l|>10^\circ$. Therefore,
there has to be another component in the plane taking its place at greater 
absolute longitudes to make the counts almost flat up to $l=27^\circ$, where it 
then stops quite suddenly.
This component  contributes around 50\% of the detected sources, so
it must be a major feature in the inner Galaxy. This component is principally 
seen at positive longitudes, extending from the bulge up to $l=27^\circ$ 
and somewhat less at negative longitudes. 
This makes the component extremely asymmetric as seen 
from the Sun, far more so than the triaxial bulge.    

Clearly, one explanation for the asymmetry in the counts would be
extinction. This has an important role as a bar would be expected to have 
dust lanes on its leading edges (Calbet el al. 1996) and  will be discussed in
Section \ref{.ext}; however, extinction can only reduce the numbers of stars,
whereas here there is clearly an extra population in the plane.

\begin{figure}
\begin{center}
\vspace{1cm}
\mbox{\epsfig{file=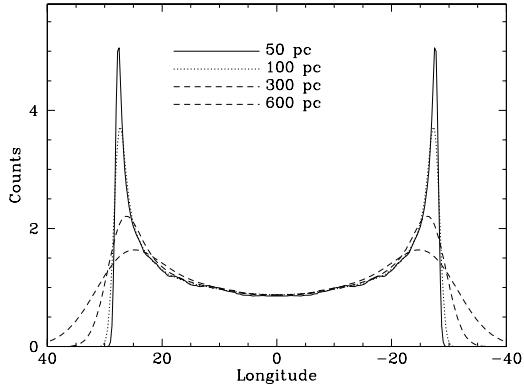,height=8cm,angle=-90}}
\end{center}
\caption{The expected shape of the counts produced by 
rings of varying thicknesses for a cut along the plane. 
The radius of the ring is 3.7~kpc and the radial 
distribution is  Gaussian with the stated sigmas. 
The counts were normalized such that each ring would give the same 
counts at $l=0^\circ$. The ring used was circular; if it were elliptical 
then the effect would be to change the position of the peaks so 
that they were no longer symmetric in longitude.}
\label{Fig:ring}
\end{figure}

In H94, a ring, by itself, is shown not to explain the in-plane
features. A ring would be expected to produce a peak in the counts at
the  positions tangential to the line of sight. If the ring is
relatively thin, then away from the tangential points the
counts should die away quickly. If the ring is very broad (e.g. as proposed
by Kent et al. 1991), then the peak becomes  a lot broader. Figure
\ref{Fig:ring} shows some possible forms that the ring counts could produce for an
axially symmetric system. If the ring were elliptical then the
longitudes of the peaks would no longer be symmetric; however, the shapes
of the peaks would remain basically unaltered. Clearly, it is possible that a
contrived ring (particularly a patchy ring), coupled with a highly
improbable distribution of extinction, could reproduce the form of the
in-plane counts. However, as will be shown, this
distribution of extinction does not exist and hence the ring by itself
cannot explain the in-plane counts.

The other alternative to explain the asymmetry in the star counts is the
existence of a bar. The work in H94 and H00 limits the possible 
orientations of the bar. The predicted bar has the near end at $l=27^\circ$ 
at a distance of 5.7 kpc and the far end at $l=-12^\circ$ at a distance of
about 11~kpc. We shall limit the discussion to whether the data presented here 
are consistent with the previous results rather than try to re-determine all 
 the parameters, as the arguments presented would be almost identical. 

Whilst simplistically one would expect the near end of the bar to give more 
counts than the far end,
this is not necessarily the case in the plane. Blitz \& Spergel (1991b) 
showed that the far end of
the bar can give a higher surface brightness than the near end in the 
plane; Unavane et al. (1998) and Unavane \& Gilmore (1998) predict 
a similar result for star counts. There are, however, a series of 
effects that have to be taken into account:

\begin{itemize}

\item The number of 
stars per unit area of sky is dependent on the volume of the bar intersected by 
the line of sight. This is dependent on the square of the distance to the bar 
at that position coupled with  sec $\alpha$, where $\alpha$ is the 
angle between 
the bar and the line of sight. The regions at negative latitudes
are more distant from the Sun than the regions on the positive side of the
bar, and the line of sight cuts the bar at a shallower angle,
both of which factors increase the counts per unit area at negative longitudes 
in comparison with  positive longitudes.

\item For the same apparent magnitude, the limiting absolute magnitude of the
sources detected at positive longitudes is fainter than that at
negative longitudes. Up to $m_K$ = +9,
 the luminosity functions will be dominated
by young stars (roughly $M_K<-7.5$ for a source at $l=-12^\circ$ and 
$M_K<-6.0$ for a source at $l=+27^\circ $, taking extinction into account). 
The luminosity function  rises steeply over this magnitude range; this means
 that
there will be far more detectable sources per unit volume at
positive longitudes than at negative longitudes.

\item The scale height of the bar is very important. H94 showed that the
scale height for the sources in the spike at $l=21^\circ$ was around 50~pc.
 This corresponds to $b=0.5^\circ$ for the near end of the bar at
$l=27^\circ$ but $b=0.25^\circ$ for the far end. Therefore, for the
$|b|<0.25^\circ$ counts, the edge of the box is a scale height at
$l=-12^\circ$, only half a scale being at $l=27^\circ$. The effect
is to reduce the ``far end''/``near end'' count ratio. Further
from the plane the ratio is decreased even further, such that in the
region $0.5^\circ <b<1^\circ $ the far end of the bar will hardly be
seen, whereas at the near end it will still be visible. If the scale
height were a number of degrees, as in the bulge, then the effect would
be small this close to the plane.

\item The far end of the bar is at twice the distance of the near end and 
hence the extinction is likely to be greater. Furthermore, a sticklike bar 
could have a dust lane on its leading edge, which would further increase the 
extinction at negative longitudes. This gives further counts for the 
positive longitudes.

\end{itemize}
\begin{figure*}
\begin{center}
\vspace{1cm}
\mbox{\epsfig{file=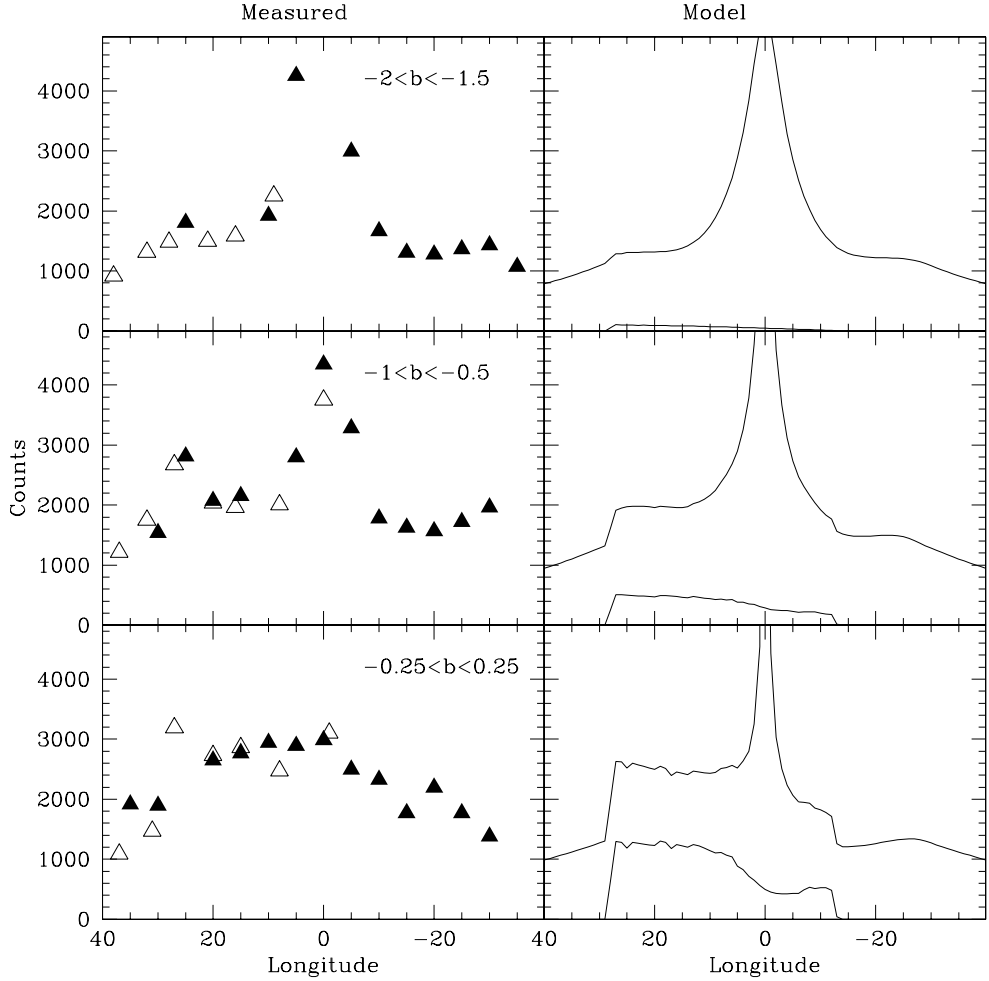,height=14cm}}
\end{center}
\caption{The left hand plots show the measured DENIS (solid triangles) 
and TMGS (open triangles) star counts to $m_{K_{\rm s}} = 9$ for the regions 
at $-2^\circ<b<-1.5^\circ$, $-1^\circ<b<-0.5^\circ$ and 
$-0.25^\circ<b<0.25^\circ$. 
The right hand plots show the prediction of the total counts of
a model containing an
inner truncated disc, bulge and 3.9 kpc bar at  position angle  
40$^\circ$. The bar contribution alone is also plotted with another solid line.}
\label{Fig:model}
\end{figure*}

Figure \ref{Fig:model} shows the measured and model counts distribution for three
latitude ranges. The model contains the truncated disc, but with the bulge and
extinction based on Wainscoat et al. (1992). The simplest possible bar 
distribution in agreement with H00 was then added. The bar used has
a thickness of 500 pc, a half-length of 4 kpc and position angle of 
43$^\circ$. The distribution was assumed to be constant along the bar but
exponential in height above the plane. The magnitude limit means that the 
sources would have to have absolute magnitudes brighter than $M_K$=-6. 
These sources will be principally young stars and so the scale height will be 
small. These sources will be the same as those in the spike seen in H94, which
have a scale height of about 50 pc. The luminosity function used was 
the same as for the disc although the density was then normalized 
to make the total counts match those at $l=27^\circ$.   

Whilst the model is simple, it does reproduce the measured counts fairly well, 
apart from the region near the GC, where the extinction model used is 
almost certainly not correct:

\begin{itemize}
\item The model predicts the sharp jump in counts near $l=27^\circ$ 
in the plane.

\item  The model predicts nearly constant counts in the plane 
between $l=27^\circ$ to about $l=-12^\circ$, at which point there is a 
drop to the disc level in the counts. 

\item In the $-1^\circ<b<-0.5^\circ$ plot, the model predicts 
that the extra sources are significant at positive longitudes but not at 
negative longitudes. The point at $l=-10^\circ$, $b=0.75^\circ$
is more or less at the level of the disc counts, whereas 
at $l=-10^\circ$ $b=0^\circ$ there is a significant extra population. 
This is caused by the small scale height of the bar.
\end{itemize}

The only major feature not predicted is the peak at $l=-22^\circ$. 
This will be dealt with later.

Unavane \& Gilmore (1998) and Unavane et al. (1998) analysed
narrow-band $L$ star counts and DENIS data at a few positions in the
Galactic plane, near the centre. Contrary to the results presented here,  they found  higher
counts at negative longitude than at positive longitude, and this fits
some of the models for the ``bar'', although, from the introduction, this
could also be classed as the triaxial bulge. By observing only a few
regions in the central few degrees, however, they were
principally measuring the  bulge, which  dominates the star counts in
the inner plane (L00). This is an old population with a scale height far
greater than the 50~pc of the feature seen here, which, as discussed
above, would increase the counts at negative longitudes. In this paper,
however,  we are dealing with a young population extending tens of
degrees along the plane from the centre. This is a very different feature, and so our
results cannot be compared with those of Unavane \& Gilmore (1998) and Unavane
et al. (1998).

The above shows that the DENIS + TMGS star counts to $m_K$=+9 between 
$l=27^\circ$ and $-12^\circ$ do have the characteristic signatures of the 
young component of a thin bar. The ring by itself, even if made elliptical, 
cannot give a good fit to the data.
Therefore, the bar proposed in H00 correctly predicts the majority 
of the features seen, without resorting to  ad hoc solutions. 

\section{Asymmetry in the extinction of the Galactic plane}
\label{.ext}

Another way of observing possible asymmetries is to examine the
distribution of  extinction along the plane.  The  proposed bar would
make the stars in the fourth quadrant significantly further away than
the bar stars in the first quadrant and, consequently, the extinction of
the bar stars would be higher at negative longitudes than at  positive
longitudes. Moreover, a possible dust lane leading the  bar (Calbet et
al. 1996) would further increase the extinction at  negative
longitudes.  For the proposed bar the extra extinction would lie between
where the dominance of the bulge in the counts  ends ($l \approx -8^\circ$) and 
somewhere near  the far tip of the bar  ($l \approx -12^\circ$).

\begin{figure}
\begin{center}
\vspace{1cm}
\mbox{\epsfig{file=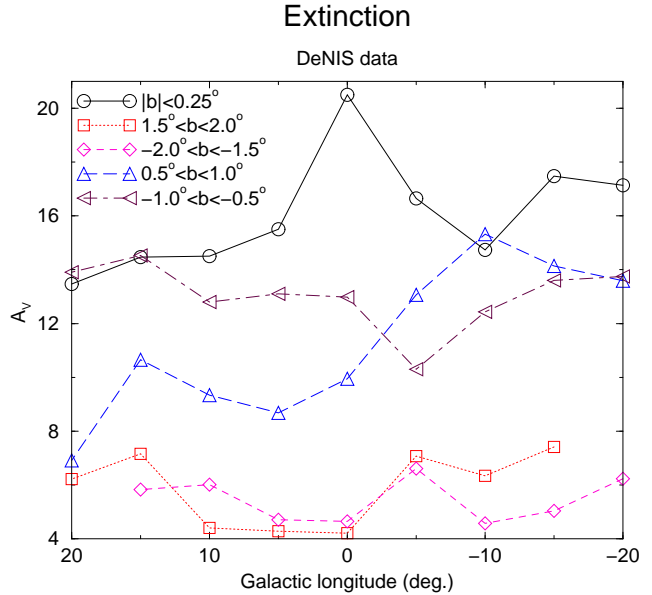,height=8cm}}
\end{center}
\caption{Plot of the extinction ($A_V$) averaged over
$\Delta l=5^\circ$ in different latitude cuts with
$\Delta b=0.5^\circ$ corresponding to the filter $V$ up to 
the furthest bulge/bar/disc stars, calculated from DENIS data by means 
of the method by Schultheis et al. (1999).}
\label{Fig:plot_av}
\end{figure}

\begin{figure*}
\begin{center}
\vspace{1cm}
\mbox{\epsfig{file=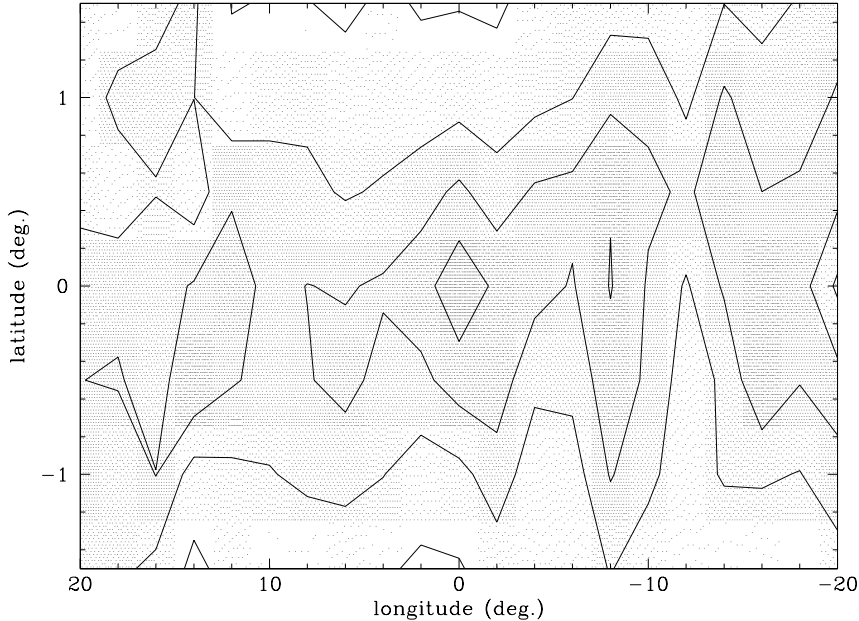,height=12cm,angle=-90}}
\end{center}
\caption{Extinction map averaged and interpolated over
$\Delta l=2^\circ$ in different latitude cuts with
$\Delta b=0.5^\circ$. Some gaps in the available data were 
substituted by interpolated data from the surroundings.
The contours stand for the extinction ($A_V$) in the filter $V$ up to 
the furthest bulge/bar/disc stars. 
This was calculated from DENIS data by means of the method of 
Schultheis et al. (1999). Contour step of 5 mag.}
\label{Fig:ext}
\end{figure*}

The calculation of the extinction is not straightforward and 
requires a number of  assumptions to be made, in particular about  the
stellar populations.  However  colour--magnitude diagrams do allow a
simple separation of disc dwarfs, disc giants and inner Galaxy giants
when there is significant extinction along the line of sight (e.g. H00
or Ruphy et al. 1997). In DENIS,  $J-K_{\rm s}$ is more appropriate since the $I$
filter does not reach the stars in the inner Galaxy along lines of
sight to regions with high extinction.  Using this idea,  Schultheis et
al. (1999) derived a method to obtain an extinction map towards the
inner Galactic plane. This  used $J$ and $K_{\rm s}$ DENIS
bands to isolate the inner Galaxy sources together with the
interstellar extinction law ($A_V$:$A_J$:$A_{K_{\rm s}}$=1:0.256:0.089) from
Glass (1999).  Theoretical isochrones (Bertelli et al. 1994) were
calculated for the RGB and AGB phase assuming a 10 Gyr population with
$Z=0.02$ and distance 8~kpc.  The $A_V$ is then determined from the
shift in the colour-magnitude diagram ($K_{\rm s}$ vs. $J-K_{\rm s}$) of the
isochrones. This method was applied in the region $|l|<8^\circ $ in the
paper by Schultheis et al. (1999). Here, the same method will be applied
for a wider region, $|l|<20^\circ $.  
The results, after smoothing  the extinction map,
are shown in Figs. \ref{Fig:plot_av} and \ref{Fig:ext}.

Asymmetry is clearly  observed in the extinction for $|b|<0.25^\circ $ and for $0.5^\circ <b<1.0^\circ $. 
In the plane there are on average two or three magnitudes more of extinction at  negative longitudes
than at positive longitudes. For off-plane regions $1.5^\circ<|b|<2.0^\circ $
the asymmetry in the extinction is not present. Again, a bar is consistent with
this result, whereas a ring is not. The extinction  calculated here is only approximate  
as it is impossible to isolate the stars of the inner Galaxy from those in the disc, 
including the disc beyond the Galactic Centre.   
However, although the exact interpretation of the value of $A_V$ is not immediately obvious, 
the asymmetry is present and this can only be
due to an intrinsic asymmetry in the central distribution of stars and/or dust.

 Fig. \ref{Fig:ext} presents the extinction as a two-dimensional map   
over the region $|l|<20^\circ $ and $|b|<1.5^\circ $.
As well as the extra extinction at negative longitudes, there is a clear
tilt in the extinction. This tilt 
 follows a path line $b\approx -0.05\times l$ for $|l|<8^\circ$, so
it runs below the plane  at positive longitudes
and above the plane at negative longitudes. 
This is a well-known feature in the CO and HI maps
(Liszt \& Burton 1980; Sanders et al. 1984). Observing a tilt
in the extinction map clearly indicates that this extinction
is due  to gas and dust in the centre of the Galaxy, where the
gas distribution is tilted. 
Because of the tilt, the asymmetry for $-1.0^\circ <|b|<-0.5^\circ $ 
is not seen in Fig. \ref{Fig:plot_av} whilst the asymmetry at
$0.5^\circ <b<1.0^\circ $ is very high. At positive longitudes, the
extinction is around 5 magnitudes less than at negative longitudes,
except for a few isolated regions, such as the excess at $(l=15^\circ,
b=-0.5^\circ)$, probably associated with a cloud in the molecular ring.

\section{An analysis of the inner Galactic plane
at  \boldmath $l<0^\circ$}

\begin{figure}
\begin{center}
\vspace{1cm}
\mbox{\epsfig{file=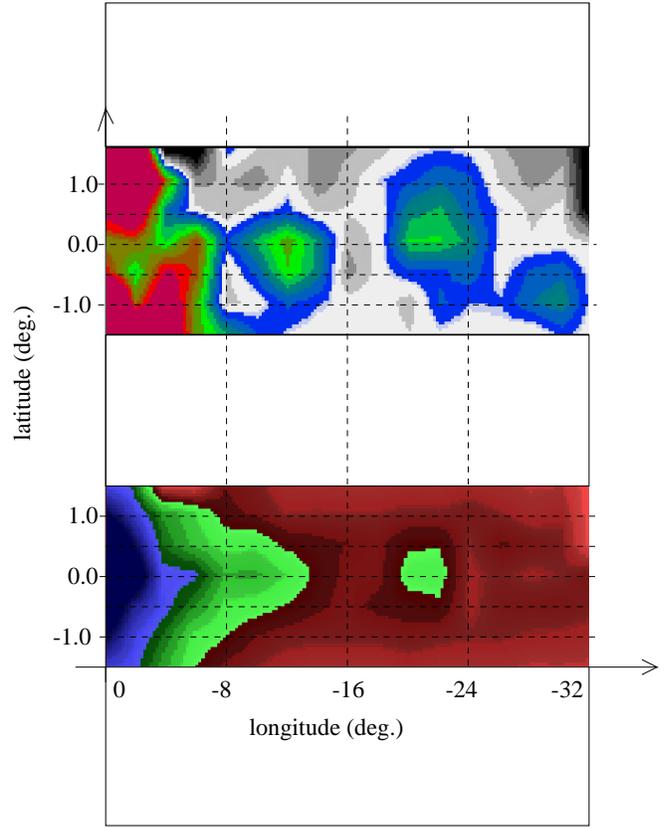,height=11cm}}
\end{center}
\caption{Map of DENIS star counts binned 
($\Delta l=2^\circ$, $\Delta b=0.5^\circ$) and interpolated. 
Region $l$ from 0$^\circ $ to $-32^\circ $; $b$ from $-1.5^\circ $
to $1.5^\circ $. Upper: $m_{K_{\rm s}}\le 9.0$; lower: 
$m_{K_{\rm s}, {\rm corrected}}\le 9.0$ and $(J-K_{\rm s})>0.5$.}
\label{Fig:plotestat2D}
\end{figure}

A map of the star counts with $m_{K_{\rm s}}\le 9.0$ 
is shown in Fig. \ref{Fig:plotestat2D} (upper) with a
binning of $\Delta l=2^\circ $ and $\Delta b=0.5^\circ $. This binning
reduces  high frequency fluctuations due to  very patchy extinction.
 Only  the negative longitudes are shown because
in this region  coverage in the DENIS batches is  almost complete  
(96 deg$^2$ are available at negative longitudes, while for positive
longitudes only 47 deg$^2$ are available). Furthermore, previous
papers have already studied the positive longitudes (e.g. H94; Kent et al. 
1991), whereas there is relatively little published
on the negative longitudes.

In the star-count map there are peaks  in a number of regions
($l=0^\circ$, $b= \pm 1.5^\circ $; $l=-5^\circ$, $b=-1^\circ $;
$l=-12^\circ$, $b=0^\circ $; $l=-22^\circ$, $b=0^\circ $;
$l=-30^\circ$, $b=-1^\circ $). Some of these peaks may be due to
real features in the star distribution, but others are clearly  due
to regions of lower extinction than in the surrounding areas. Often,  local
maxima in the star counts are present where there are local minima 
in the absorption (Fig. \ref{Fig:ext}). The extreme patchiness of the 
extinction, particularly when looking towards the inner Galaxy, means that care 
needs to be taken when interpreting counts in a few small areas, even 
when working at longer wavelengths, for example the $L$ band or 12 $\mu $m. 
The extinction can vary by 5 to 10 magnitudes in the visible 
over a few degrees, which will change $A_L$ by 0.25 to 0.5 magnitudes 
and hence will have a significant effect on the counts.  

One method of distinguishing the peaks caused by areas of lower extinction 
from real features is to derive  star counts corrected
for extinction. The measured  colour of a source is dependent on its intrinsic
colour and the extinction. When looking at a restricted apparent magnitude 
range the majority of the sources at the same distance will have the same intrinsic 
colour. Hence when the majority of the sources along a particular line of sight are at the same distance,   the extinction  to each source---and hence the magnitude that the source would  have without extinction---can be calculated as:

\[
m_{K_{\rm s}, {\rm corrected}}=m_{K_{\rm s}}-\frac{A_{K_{\rm s}}}{A_J-A_{K_{\rm s}}}\lbrack(J-K_{\rm s})-1.0\rbrack
\]
if
\[
(J-K_{\rm s})>1.0.\]
and
\[
\frac{A_{K_{\rm s}}}{A_J-A_{K_{\rm s}}}=\frac{3}{5}.
\]

The correction is calculated from the difference between the measured 
$(J-K_{\rm s})$ and the expected 
intrinsic colour of the sources. At the distance of the bulge,
 $m_K$ = +9 means that the sources are early M-giants,
 and so the intrinsic colour
will be about 1.1. The procedure that we have implemented only dereddens sources with $J-K>1.0$, so that only the  highly extinguished sources  will be corrected. Furthermore, in order to remove many of the 
local disc sources and so improve the contrast, all sources with  
$(J-K_{\rm s})<0.5$  were  removed.
This means that  the inner Galaxy sources
will be  preferentially  corrected, whereas the local disc sources, which
are not as reddened, will not be corrected or removed.  
This approach is roughly consistent 
with that of Schultheis et al. (1999), except that here each star is 
individually corrected for extinction before each area is averaged. 
 The star counts corrected
for extinction using this method  are shown in the lower plot of  Figure 
\ref{Fig:plotestat2D}.

This approach is
reasonable  because of the very high density of sources in the inner
Galaxy. In many regions over 50\% of the sources  at a particular
magnitude are from the inner Galaxy and so by improving the ratio the vast
majority of the detected sources come from a  relatively restricted
distance range (8 $\pm$ 3~kpc).  To have  an extinction-corrected apparent
$K$ magnitude of +9, the sources would need to have an absolute magnitude
of between $-4.5$ and $-6.2$, which would mean that the typical intrinsic 
$J-K$ colour is around 1.0. 
Even including the more luminous sources, the intrinsic $J-K$ 
will typically be  at most 1.3. 
An error of $\sim 0.2$ magnitudes 
in the assumed intrinsic colour leads to an error in the
 extinction-corrected $K_{\rm s}$ magnitude of  about 0.1 magnitudes, which is 
negligible. Clearly some disc sources will be included and these will end up 
with incorrect magnitudes. However, this will be a small proportion of the 
total sources and their distribution is symmetrical and in principle 
predictable, so  at most there will be a loss of contrast for the inner 
Galaxy features. Hence for the case where the majority of the 
sources are relatively concentrated in a certain location along the line of 
sight this is a straightforward method for recovering the form of the 
underlying star distribution, at least in the infrared. 

It should be noted that this method needs to detect the source in both
$J$ and $K_{\rm s}$, and so can only remove the effect of extinction up to a
certain extinction limit. When the extinction becomes too high, the
source will not be detected at $J$ and so sources will be lost. The  $J$
limiting magnitude in DENIS is +16.5 and so the extinction correction
limit when making $K$ counts to +9 will be about 20 magnitudes in $V$.  The
nuclei of the dense dust clouds, particularly near the Galactic Centre,
will have far more extinction than this and in these regions hardly any
sources will be detected. In general, these regions are small and so when
averaging the counts over a reasonable area the losses will be small.
However, the extinction-corrected counts shown here should be examined
qualitatively, not quantitatively.

The most noticeable effect of the extinction-corrected counts is that a
plot that was very patchy is converted into one with a basically 
smooth distribution of sources. Most of the local troughs have been
``filled in'' and the peaks disappear. 
It should also be noted that, unlike the extinction map, the stellar
distribution shows no tilt. This may be important for the study of the 
causes of disc warps and inner tilts in the galaxies, 
which may be related (Ostriker \& Binney 1989). 
A force that distorts the gas rather than the stars may indicate
an origin related to magnetic fields (Porcel et al. 1997) or
accretion of the intergalactic medium (L\'opez-Corredoira et al. 2001b).
Another possibility is that as the inner disc is truncated 
there would be few stars in the inner disc to be detected, 
and so even if there were a tilt it would not be seen.

The results of Fig. \ref{Fig:plotestat2D} suggest that the extinction in Fig.
\ref{Fig:ext} is consistent with that derived by the colour-term
correction. Hence, by comparing the different maps the following
conclusion about the peaks and valleys in Fig. \ref{Fig:plotestat2D}
can be made:

\begin{description}
\item[$l=0^\circ$, $b=0^\circ $]: In the uncorrected counts there was a 
valley with peaks on either side ($b=\pm 1.5^\circ$). The maximum source density is 
expected to be at $l=0^\circ$, $b=0^\circ $, but there is too much extinction in this
region (over $A_V\approx 20$ from  Fig. \ref{Fig:ext}).  When 
corrected  for  extinction, the two peaks are substituted for the one at
the centre of the Galaxy.

\item[$l=-5^\circ$, $b=-1^\circ $]: In the original counts this local maximum 
in the star counts coincides with a local minimum in extinction (Fig. \ref{Fig:ext}). 
When the counts are corrected for extinction, counts from the neighbouring 
regions increase and the peak disappears. This confirms that 
this peak is caused by a line of sight which does not pass through a major 
extinction region.

\item[$l=-8^\circ$ to $l=-10^\circ$, $b=0^\circ $]: This is a region of high 
extinction and a dip in the  counts. When  corrected for extinction the  
counts recover, the dip  disappears and this region becomes part 
of the spur extending to about $l=-14^\circ$.

\item[$l=-12^\circ$ to $l=-14^\circ$, $b=0^\circ $]: The absorption map
in Fig. \ref{Fig:ext} shows that the extinction is lower than in the
neighbouring areas and hence that there is a peak in the counts at this
location. We suspect that the peak of old OH/IR stars at $l=-10^\circ $ described 
by Sevenster (1999) has something to do with the present peak at $l=-12^\circ $
for young stars. When the counts are correct for extinction the peak at
$l=-12^\circ$ is converted into a ``spur'' in the counts 
jutting out about 5 degrees
from the bulge but only existing very close to the plane. This spur
ends at about $l=-14^\circ$. It is in the stellar distribution
jutting out from the bulge, and fits precisely the far end of the bar
described in H00.

\item[$l=-15^\circ$ to $l=-19^\circ$, $b=0^\circ $]: This region has
significantly lower counts than the neighbouring regions before
correcting for extinction, which in this region is apparently high. 
Calbet  et al. (1996)  also show this to be a region of
high extinction from their analysis of the {\em COBE}/DIRBE surface brightness
maps. After correcting for the extinction, the counts do not
recover to the same level as the neighbouring regions. This  could be
explained if this were a  a zone of extremely  high
extinction ($A_V>30$ mag); however, looking in detail  at the distribution
of the sources,  there is no evidence for the dark ``shadows'' caused by
 dense molecular clouds, such as those that are seen towards
the Galactic Centre, and so this is unlikely.  A more realistic
solution is that there are in fact few inner Galaxy sources along these
lines of sight. In all lines of sight, sources will be detected from
the far side of the disc.  These are at a great distance and so are highly
reddened. For most lines of sight towards the inner Galaxy, the inner
Galaxy sources are by far the most numerous, so they dominate in any
analysis. However, if these source are not present then the far side of
the disc will become more important and make the average extinction
higher when determined by the methods used here.

\item[$l=-20^\circ$ to $l=-24^\circ$, $b=0^\circ $]: This maximum
extends over a number of  degrees in the plane and is still 
present  when the magnitudes are corrected for extinction; hence, this
is clearly a real feature in the star distribution and not an artefact
caused by patchy extinction in the nearby regions. In H94 this region
was proposed to be the far end of the bar.  However, this was based on
an analysis of the {\em COBE}/DIRBE surface brightness maps, 
which are inherently ambiguous as these offer no information on distance. 
The data presented in H00 and the new data presented here both show  
that if there is a bar then the far end is near $l=-12^\circ$. The peak, centred at
$l=-22^\circ$, has more or less the form  that is expected for the
tangential point to a ring or spiral arm. A full discussion on this is
given in the following section.

\item[$l=-30^\circ$, $b=-1^\circ $]: This peak almost disappears
when the extinction correction is carried out, hence it is just an area of lower extinction. 
\end{description}

\section{Discussion}

\subsection{The ring}
The ring by itself cannot explain the form of the counts along the
Galactic plane. However this does not mean that there cannot be a
ring. More than three quarters of known
barred galaxies have rings (Buta 1996), so if
the Galaxy is barred then there is a high chance that it will
have a ring as well. The
prominence of the  peak at $l=-22^\circ$ indicates that it is caused by
an important feature in the inner Galaxy.  Its form indicates that it
is the tangential point to a ring or spiral arm, and that its location
coincides with tangent from the ``3 kpc arm'' seen in the CO maps.
The 3 kpc arm is an unusual radio feature in the plane with a radial 
velocity of $-53$ km s$^{-1}$ at $l=0^\circ$ rather than approximately zero as for
the other arms. Furthermore, the tangential point to the 
Norma arm is at about $l=-30^\circ$, yet there is no clear evidence 
for the arm in the counts. The most likely
explanation for the peak at $l=-22^\circ$ is that it is 
the tangential point to a ring, or more probably a pseudo-ring
(i.e. the inner arms are very tightly wound around the bar to form what
is almost a ring, as in NGC 1433). Sevenster (1999)
detected an excess of OH/IR stars at $l=-22^\circ $ and suggested that
the 3 kpc arm was the cause. He also concluded that this feature is
indeed ringlike.

At $l=-22^\circ$ the radio tangential point and star-count peaks almost coincide, whereas 
at tangential points at positive longitudes in radio, $l=24^\circ$, 
there is actually  a strong dip in the counts. 
The closest peak to $l=24^\circ$ is at $l=27^\circ$, hence it is
probable that the tangential point to the ring coincides with the end
of the bar, which produces the strong peak in the counts seen  
at the Galactic Centre itself. The reason that the stars and gas are
somewhat separated at positive longitudes whilst at negative
longitudes they are close together is probably because at positive
longitudes the tangential point to the ring almost coincides with the end of 
the bar. The ring is elliptical and, assuming that the main
axis were parallel to the bar, which is not necessarily the case
(Sandage \& Bedke 1994), the axial ratio of the ring would be
1:0.76 for the stars and some what higher for the gas. This would compare
with the mean axial ratio for rings of 0.81 $\pm$ 0.06 (Buta
1996). The diameter of the ring would be the same as the bar, hence 8 kpc,
again very close to the typical values of 9 kpc found in external Galaxies 
(Freeman 1996).

\subsection{Parameters of the bar}
Combining the results from this paper and those presented in H00 and H94
the principle parameters of the bar can be determined. 

\begin{description}

\item[The position of the ends of the bar]: 
The near end of the bar is in the first quadrant at $l=27^\circ$. 
At this position there is a sudden change in the stellar density and luminosity 
functions. The far end of the bar is in the third quadrant and is seen as the 
spur extending from the bulge to $l=-14^\circ $.

\item[Position angle:]  Using the above positions and assuming that the 
bar is rectilinear, centred at the Galactic centre and with a  width of 
500 pc, this gives a position angle between the bar 
and the solar Galactic Centre radius vector of around 40 $\pm 5$ deg.  
The error comes from the uncertainty in determining the position of the 
ends but the result is consistent with the other determinations for the 
in-plane bar (e.g. H00, or Weinberg 1992).

\item[Distance to the bar:] The distance to the bar varies from 
$d=5.7$ kpc at the closest position ($l=27^\circ $) to $d=11.1$ kpc
at the farthest position ($l=-14^\circ $).  

\item[Length of the bar:] The distance from the centre of the Galaxy 
to the end  of the bar is $\sim 0.48R_0$, or $\sim 3.9$ kpc 
(the distance to the Galactic Centre is taken as $R_0=7.9$ kpc; 
L00).

\item[Vertical thickness:] Figure \ref{Fig:DENIScounts9} shows that the 
counts are very asymmetric at  $|b|<0.25^\circ $, somewhat less  so at 
$0.5^\circ <|b|<1.0^\circ$, but symmetric
 by  $1.5^\circ <|b|<2.0^\circ$.
At $|b|<0.25^\circ $ there is an average gradient between $l=25^\circ$ and 
$l=-25^\circ$ of  $\sim 30$ (star/deg$^2$)/deg, which drops to  $\sim 20$ 
(star/deg$^2$)/deg at $0.5^\circ <|b|<1.0^\circ$, but is only   
$\sim 3$ (star/deg$^2$)/deg) or less at $1.5^\circ <|b|<2.0^\circ$.
The projected vertical FWHM of the
bar is therefore around 1 deg for stars with $M_{K_{\rm s}}<-6.0$ 
(or $M_{K_{\rm s}}<-7.5$
in the farthest positions of the bar). This implies that the scale height 
of the young stellar population of the bar discussed here is about 50 pc.
However, it should be noted that the older component detected  
in H00 will probably have a significantly larger scale height.

\item[Star density:] From  Figure \ref{Fig:DENIScounts9} for
$|b|<0.25^\circ $, there is  an excess of $\sim 1500$ stars/deg$^2$
over the disc at $l\approx 20^\circ$ when compared to $l\approx 30^\circ $,
and this excess will be mainly due to the bar stars at that location. 
The extinction at this position is $A_V\approx 13$ mag 
(see Fig. \ref{Fig:plot_av}), and the distance of the bar stars is
$d\approx 6$ kpc.
Therefore, assuming a bar width of $\sim 500$ pc,
the density of bar stars up to $m_{K_{\rm s}}=9.0$ is 
$\rho \sim 3\times 10^{-4}$ star/pc$^3$,
i.e.  density of bar stars brighter than 
$M_{K_{\rm s}}=-6.0$ (mainly late giants, supergiants and a few AGBs).
For comparison, the bulge has the same density of stars
at about 700 pc from the Galactic Centre in the plane
(L00).

\end{description}

\section{Conclusions}

The following features are seen in the plane in the inner  
regions of the Galaxy ($|l|<30^\circ $):

\begin{itemize}

\item  The star counts are very asymmetric with longitude. 
There is a major excess of star counts at positive 
with respect to negative longitudes in the Galactic plane ($|b|<1^\circ $).

\item  On average, there is more extinction at negative 
than at  positive longitudes in the Galactic plane ($|b|<1^\circ $).

\item  Off-plane star counts  ($1.5^\circ <|b|<2.0^\circ $) are symmetric in longitude and  are nearly constant for  $12^\circ <|l|<30^\circ$.

\item  There is a peak in the stellar distribution near  $l=-22^\circ $,
$b=0^\circ $, that is not due to a hole in the extinction.

\item  There is a spur in the counts near the plane extending from the bulge 
to  $l\approx -14^\circ $.

\item There is a tilt of the gas in the inner disc that is not  seen in
the stellar component of the central Galaxy.

\end{itemize}

The asymmetry in the in-plane 2.2 $\mu$m star counts 
for $30^\circ>l>-30^\circ$ cannot be caused by the bulge, and 
other solutions  such as a ring or bar by themselves, spiral arms, etc., 
simply do not work.  Only a ringed barred spiral with an inner truncated 
disc will naturally reproduce the features seen.

The parameters determined  
for the bar and ring make the Milky Way  a fairly typical barred 
galaxy. We agree with Sevenster (1999) that NGC 1433 or M95 is a reasonable 
approximation of what the Galaxy would look like if viewed face on (Fig.
\ref{Fig:M95}, \ref{Fig:NGC1433}) giving a revised Hubble type
(Sandage \& Tammann 1981) of SBb(s) I--II. 
The closest part of the bar is in the first quadrant at $l=27^\circ $
and the tip at negative Galactic latitudes is near $l=-14^\circ $.
The bar runs from the bulge  to 3.9 kpc and is  
as luminous as the bulge at about 700 pc from the 
GC,and  contains both young and old stars.
The inner disc is truncated, with the truncation starting near 
the radius of the ring. 
It is also proposed that the feature at $l=-22^\circ$ is the tangential point to  stellar ring. 
The ring is elliptical and  runs through the end 
of the bar. The spiral arms are probably far weaker than the ring. 

Clearly a significant amount of work remains to be done
to tie down the parameters with more precision. 
However, the information now available, particularly from the 2.2 micron 
star counts, does allow the tentative identification of the Milky Way as an 
early-type ringed barred Galaxy with a Freeman type II disc.   

\vspace{10mm}

{\bf Acknowledgements:}
We gratefully acknowledge the anonymous referee for helpful comments.
DENIS is the result of a joint effort involving personnel and financial
contributions from several Institutes, mostly located in Europe,
including the the Instituto de Astrof\'{\i}sica de Canarias (IAC). It has been
supported financially mainly by the French Institute National des Sciences de
l'Univers, CNRS, and French Education Ministry, the European Southern 
Observatory, the State of Baden-W\"urttemberg, and the European Commission 
under a network of the Human Capital and Mobility programme.
TMGS observations were made at the
Carlos S\'anchez Telescope, which is operated by the IAC at its
Observatorio del Teide on the island of Tenerife.

%\eject

%{\bf Figure captions}

%\begin{description}

%\item[Figure 1:]

%\end{description}

\end{document}